\documentclass[11pt]{article}
\usepackage{bbm}
\usepackage{amscd}
\usepackage{amsfonts}
\usepackage{CJK}
\usepackage{graphicx}
\usepackage{amssymb}
\usepackage{amsmath}
\usepackage{color}
\usepackage{cite}

\newcommand\btd{\raise 2pt \hbox{$\hat\bigtriangledown$}\hskip 1.5pt}
\newcommand\bt{\raise 2pt \hbox{$\bigtriangledown$}\hskip 1.5pt}

\def\no{\nonumber}

\begin{document}
\title{Approximate homotopy series solutions of perturbed PDEs via approximate symmetry method}
\author{Zhiyong Zhang \footnote{  E-mail: zhiyong-2008@163.com} 
\\\small KLMM, Institute of Systems Science, AMSS, Chinese Academy of Sciences,\\
\small Beijing 100190,  P.R. China\\
}
\date{}
\maketitle

\noindent{\bf Abstract:}  We show that the two couple equations
derived by approximate symmetry method and approximate homotopy
symmetry method are connected by a transformation for the perturbed
PDEs. Consequently, approximate homotopy series solutions can be
obtained by acting the transformation on the known solutions by
approximate symmetry method. Applications  to the Cahn-Hilliard
equation illustrate the effectiveness of the transformation.



\section{Introduction}
Group theoretical methods based on local and nonlocal symmetries
provide remarkable  techniques to analyze partial differential
equations (PDEs) such as constructing group invariant solutions,
studying integrability etc., and also build a ladder to explore
profound understandings of the physics of the underlying problems
such as detecting conservation laws \cite{olv,blu1}. The
prerequisite of achieving these goals is that the PDEs under study
possess affluent symmetries, nevertheless, for the perturbed PDEs
(containing a small perturbed parameter), the classical symmetry
method losses its superiorities because the small parameter disturbs
the symmetry group properties of the unperturbed equation, even
makes the studied PDEs hold few symmetries.

In view of the drawbacks of classical symmetry method for conducting
this type of equations, the integration of perturbation analysis and
symmetry group theory makes two well-known approximate symmetry
methods (ASM) emerge \cite{ww,ba}. The originators of the two
approaches are Baikov \emph{et.al.} \cite{ba} and Fushchich and
Shtelen \cite{ww} who employ the standard perturbation techniques on
the infinitesimal operators and dependent variables respectively.
The comparisons of the two ASM which are illustrated by several
specific equations may be found in \cite{pak,ron}. Quite recently,
the extended combination of homotopy conception with perturbation
technique and symmetry method produces approximate homotopy symmetry
method (AHSM), which is suitable for studying the problems which
don't contain small parameters \cite{jiao}.

The distinct merit of AHSM  is that the accuracy of approximate homotopy series
solutions can be controlled by adjusting the convergence-control
parameter, however, it is at the expense of the greater computation
than ASM \cite{zhang}. On the other hand, we find that the frameworks
of ASM and AHSM for the perturbed PDEs are quite similar except for
the different studied objects
(see subsection 2.1 below). 
Hence, a natural question occurs: what is the relationship between
ASM and AHSM, does there exist a transformation or others to link
them? In \cite{zhang}, the authors study this problem which is
exemplified by a class of perturbed nonlinear wave equations and
show that first-order coupled equations and higher-order similarity
reduced equations are connected by two scaling transformations
respectively. However, the general results for the connections
between ASM and AHSM for general perturbed PDEs are still open.

The purpose of our paper is  to give a positive answer about this
question. We show that the two couple equations obtained by ASM and
AHSM are equivalent under a transformation. Based on this
transformation, one can construct approximate homotopy series
solutions directly from the known approximate solutions by ASM
without performing AHSM for the governing equations.

The outline of the paper is as follows: In section 2, we briefly
review some basic notions and give the main results. Section 3
concentrate on the applications of the results to the Cahn-Hilliard
equation. The last section contains conclusion and discussion of our
results.

\section{Basic notions and principles}
In this section, we first briefly review the main ideas of ASM and
AHSM, and then give the results about the relationship between the
two approaches.
\subsection{Basic notions}
We take the following $k$th-order  perturbed nonlinear PDE
\begin{eqnarray}\label{equation}
E=E_0(x,u,u^{(1)},\dots,u^{(k)})+\epsilon\,
E_1(x,u,u^{(1)},\dots,u^{(k)})=0
\end{eqnarray}
into account to illustrate the main ideas of ASM and AHSM,
throughout the section, with independent variables $x=(x_1, \dots,
x_n)(n\geq 2)$, a single dependent variable $u$, $u^{(j)}=\partial^j
u/\partial x_1^{j_1}\dots\partial x_n^{j_n}$ with
$j_1+\dots+j_n=j~(j,j_1,\dots,j_n = 1, \dots , k)$ is a collection
of all $j$th-order partial derivatives, $\epsilon$ is a small
perturbed parameter. Note that we only use one variable $u$ to keep
the notation simple, the generalization to systems being fairly
obvious. Also $E_i=E_i(x,u,u^{(1)},\dots,u^{(k)})(i=0,1)$ if no
special notes are added.

Let's first recall the definition of classical symmetry. Consider
the infinitesimal generator
\begin{eqnarray}\label{oper}
X=\sum_{i=1}^n\xi^i(x,u_0,u_1\dots,u^{(k)})\partial_{x_i}
+\eta(x,u_0,u_1\dots,u^{(k)})\partial_u,
\end{eqnarray}
where $\xi^i$ and $\eta$ are smooth functions of their arguments.
Let $X^{(k)}$ be the $k$th-prolongation of the generator $X$
calculated by the well-known prolongation formulae \cite{olv,blu1}.

 \textbf{Definition 1.} (Symmetry or
exact symmetry \cite{olv,blu1}) The Lie symmetry $X$ of the form
(\ref{oper}) is admitted by Eq.(\ref{equation}) if and only if
\begin{eqnarray}\label{det}
&&\no X^{(k)}E|_{E=0}=0,
\end{eqnarray}
where the notation $|_{E=0}$ means evaluated on the solution
manifold of $E=0$.

The ASM originated by Fushchich and Shtelen employs a perturbation
of dependent variable and then approximate symmetry of original
equation is defined to the exact symmetry of the system
corresponding to each order in the small parameter.

Specifically, expanding the dependent variable with respect to the
small parameter $\epsilon$ yields
\begin{eqnarray}\label{tran}
&&\no u =\sum^{\infty}_{l=0}\epsilon^l u_l,
\end{eqnarray}
where $u_l=u_l(x)\,(l=0,1,2,\dots)$ stand for unknown functions
throughout the paper, then substituting it into Eq.(\ref{equation})
and vanishing the coefficients of all different powers of
$\epsilon$, one obtains a coupled system
\begin{eqnarray}\label{approdeter}
&& \no \epsilon^0:~~E_0|_{\epsilon=0}=0,\\
&& \epsilon^i:~~\Big[\frac{\partial^i E_0}{\partial
\epsilon^i}+i\,\frac{\partial^{i-1} E_1}{\partial
\epsilon^{i-1}}\Big]_{|_{\epsilon=0}}=0,\quad i=1,2,\dots.
\end{eqnarray}
 \textbf{Definition
2.} (Approximate symmetry \cite{ww}) The $i$th order approximate
symmetry of Eq.(\ref{equation}) is defined to the exact symmetry of
 the first $i+1$ equations  in Eq.(\ref{approdeter}).

For approximate homotopy symmetry of Eq.(\ref{equation}), one
generally consider the following simple homotopy
model\begin{eqnarray}\label{homotopymodel}
 && H(u,q)=(1-\theta q)E_0+q\epsilon (1-\theta) E_1=0,
\end{eqnarray}
where, hereinafter, $q\in [ 0,1]$ is an embedding homotopy parameter
and $\theta$ denotes the convergence-control parameter. The above
homotopy model has the property $H(u,0)=E_0,\, H(u,1)=E$.

Assuming Eq.(\ref{equation}) has the  homotopy series solutions in
the form
\begin{eqnarray}\label{tranhomo}
&& u =\sum^{\infty}_{l=0}q^l u_l,
\end{eqnarray}
where $u_l$ solve
\begin{eqnarray}\label{homotopydeter}
&& \no q^0:~~ E_0|_{q=0}=0,\\
&&q^i: ~~\Big[\frac{\partial^i E_0}{\partial q^i}-i\theta
\frac{\partial^{i-1} E_0}{\partial
q^{i-1}}+i\epsilon(1-\theta)\frac{\partial^{i-1} E_1}{\partial
q^{i-1}}\Big]_{|_{q=0}}=0,~~~i=1,2,\dots.
\end{eqnarray}

Equivalently, we rearrange Eq.(\ref{homotopydeter}) by eliminating
$\partial^{i-1} E_0/\partial q^{i-1}$ in the $(i+1)$th equation by
means of the first $i$ equations and transform it into the following
form
\begin{eqnarray}\label{homotopydetert}
&& \no q^0:~~ E_0|_{q=0}=0,\\
&&q^i: ~~\Big[\frac{\partial^i E_0}{\partial
q^i}+\epsilon(1-\theta)\sum_{k=0}^{i-1}\theta^{i-1-k}\frac{i!}{k!}\frac{\partial^{k}
E_1}{\partial q^{k}}\Big]_{|_{q=0}}=0,~~~i=1,2,\dots.
\end{eqnarray}

 \textbf{Definition 3.} (Approximate
homotopy symmetry \cite{jiao}) The $i$th  order approximate homotopy
symmetry of Eq.(\ref{equation}) associated with homotopy model
(\ref{homotopymodel}) corresponds to the exact symmetry of the first
$i + 1$ equations  in Eq.(\ref{homotopydeter}) or
Eq.(\ref{homotopydetert}).

Obviously, the first equation of Eq.(\ref{approdeter}) is the same
as the first one of Eq.(\ref{homotopydetert}). When $\theta=0$,
Eq.(\ref{approdeter}) is equivalent to Eq.(\ref{homotopydetert})
under the scaling transformation $u_i=\epsilon^i \widetilde{u}_i
~(i=0,1,2,\dots)$ owing to $(\partial^{i} E_0/\partial
q^{i})_{|_{q=0}}$ is linear about $u_i$ and its derivatives
$u_i^{(\lambda)}$ (see Lemma 1 below). However, the parameter
$\theta$ is used to adjust the convergence of homotopy series
solution and may not be zero, thus it is the intention below to
search for the connections under the condition $\theta\neq0$.
\subsection{Main results}
In what follows, we first prove two basic lemmas and then establish
the connections between Eq.(\ref{approdeter}) and
Eq.(\ref{homotopydetert}).

 \textbf{Lemma 1.} $(\partial^{n} E_0/\partial q^{n})_{|_{q=0}}$ is
 linear about $u_{n}$ and its derivatives  $u_n^{(\lambda)}$, where $n(\geq1)$ are positive
 integers  and $ \lambda=0,1,\dots,k$.

\emph{ Proof:} 
We take $u^{(0)}(=u),u^{(1)},\dots,u^{(k)}$ of $E_0$ as independent
arguments and expand $u$ as (\ref{tranhomo}),
then by means of the generalized Faa di bruno's formula \cite{fb},
$n$th-order derivative of $E_0$ with respect to $q$ is
\begin{eqnarray}\label{total1}
&&\no \frac{\partial^{n} E_0}{\partial q^{n}}
=\sum_0\sum_1\dots\sum_n\frac{n!}{\prod\limits_{i=1}^n(i!)^{r_i}\prod\limits_{i=1}^n
\prod\limits_{j=1}^{k+1}p_{ij}!}\times\frac{\partial^r E_0}{\partial
[u^{(0)}]^{p_1}\partial [u^{(1)}]^{p_2}\dots
\partial [u^{(k)}]^{p_{k+1}}}\\&&\hspace{1.6cm}\times\prod_{i=1}^n
\left(\frac{\partial^i u^{(0)}}{\partial q^i}\right)^{p_{i1}}
\left(\frac{\partial^i u^{(1)}}{\partial q^i}\right)^{p_{i2}}\dots
\left(\frac{\partial^i u^{(k)}}{\partial q^i}\right)^{p_{i(k+1)}},
\end{eqnarray}
where the respective sums are over all nonnegative integer solutions of the Diophantine
equations, as follows
\begin{eqnarray}\label{para}
&&\no \sum_0: ~r_1+2r_2+\dots+n\,r_n=n,
\\\no&&\sum_m:~ p_{m1}+p_{m2}
+\dots+p_{m(k+1)}=r_m,~~~~m=1,2,\dots, n,
\\\no&&r=p_1+p_2+\dots+p_{k+1}=r_1+r_2+\dots+r_n,\\&&p_j=p_{1j}+p_{2j}+\dots+p_{nj},~~~~j=1,2,\dots,k+1,
\end{eqnarray}
and
\begin{eqnarray}\label{derivative}
\frac{\partial^{i}u^{(\lambda)}}{\partial
q^{i}}=\sum^{\infty}_{l=i}\frac{l!}{(l-i)!}~q^{l-i}
u_{l}^{(\lambda)},~~~\lambda=0,1,\dots,k.
\end{eqnarray}

Obviously, when $q=0$, all terms in Eq.(\ref{derivative}) vanish
except for $u_i^{(\lambda)}$.  Similarly, Eq.(\ref{total1}) with
$q=0$ becomes
\begin{eqnarray}\label{total2}
&&\no \frac{\partial^{n} E_0}{\partial q^{n}}_{|_{q=0}}
=\sum_0\sum_1\dots\sum_n\frac{n!}{\prod\limits_{i=1}^n
\prod\limits_{j=1}^{k+1}p_{ij}!}\times\left[\frac{\partial^r
E_0}{\partial [u^{(0)}]^{p_1}\partial [u^{(1)}]^{p_2}\dots
\partial [u^{(k)}]^{p_{k+1}}}\right]_{|_{q=0}}\\&&\hspace{2.2cm}\times\prod_{i=1}^n
\left[u_i^{(0)}\right] ^{p_{i1}}\left[u_i^{(1)}\right]
^{p_{i2}}\dots \left[u_i^{(k)}\right] ^{p_{i(k+1)}}.
\end{eqnarray}

In Eq.(\ref{total2}), only when $i=n$, the terms containing $u_n$(or
its derivatives) occur. At the moment, the solution of the first
equation in Eq.(\ref{para}) $r_n=1,r_i=0(i\neq n)$ generates the
terms uniquely, then other equations give $r=1$ and $p_i=1, p_s=0
(s\neq i)$ and $p_{ni}=1, p_{ns}=0 (s\neq i)$. Hence, the arguments
$u_n$ and its derivatives only occur in the following terms in
Eq.(\ref{total2})
\begin{eqnarray}
&&\no\frac{\partial E_0}{\partial u^{(0)}}_{|_{q=0}}u^{(0)}_n,~~
\frac{\partial E_0}{\partial u^{(1)}}_{|_{q=0}}u^{(1)}_n,~~\dots,~~
\frac{\partial E_0}{\partial u^{(k)}}_{|_{q=0}}u^{(k)}_n,
\end{eqnarray}
which are linear about  $u_n$ and its derivatives since $(\partial
E_0/\partial u^{(i)})_{|_{q=0}}\,(i=0,1,\dots,k)$ are only functions
of $x,u_0$ and their derivatives.
 Meanwhile, all other terms with $q=0$ are only functions of
$x,u,u_1,\dots,u_{n-1}$ and their derivatives, thus 
Lemma 1 follows. The proof ends.$~~~~\square$

\textbf{Lemma 2.} Substituting the transformation
\begin{eqnarray}\label{tranY22}
&&
u_0^{(\lambda)}=\widehat{u}_0^{(\lambda)},~~~~u_l^{(\lambda)}=\sum^{l-1}_{j=0}{l-1\choose
j}\theta^j\widehat{u}^{(\lambda)}_{l-j},~~l=1,2,\dots,~\lambda=0,1,\dots,k,
\end{eqnarray}
into $(\partial^n E_0/\partial q^n)_{|_{q=0}}$, after deleting the
symbol $\,\widehat{}\,$,  $[\partial^n E_0(\theta:
u,\dots,u^{(k)})/\partial q^n]_{|_{q=0}}$ is equal to
\begin{eqnarray}
\Big[\frac{\partial^n E_0}{\partial
q^n}+\sum\limits_{i=2}^{n}\frac{n!}{(i-1)!}{n-1\choose
i-2}\theta^{n-i+1}\frac{\partial^{i-1} E_0}{\partial
q^{i-1}}\Big]_{|_{q=0}},
\end{eqnarray}
where $[\partial^n E_0(\theta: u,\dots,u^{(k)})/\partial
q^n]_{|_{q=0}}$ denotes the transformed expression via
(\ref{tranY22}) and \\$(\partial^i E_0/\partial
q^i)_{|_{q=0}}(i=2,3,\dots, n)$ stand for the untransformed ones.

\emph{Proof:} The main idea of the proof is to regard $[\partial^n
E_0(\theta: u,\dots,u^{(k)})/\partial q^n]_{|_{q=0}}$ as a
polynomial of $\theta$ and search for the coefficients of different
degree of $\theta$ in it.


Specifically, transformation (\ref{tranY22}) makes Eq.(\ref{total2})
become
\begin{eqnarray}\label{total3}
&&\no \frac{\partial^{n} E_0}{\partial q^{n}}_{|_{q=0}}
=\sum_0\sum_1\dots\sum_n\frac{n!}{\prod\limits_{i=1}^n
\prod\limits_{j=1}^{k+1}p_{ij}!}\times\left[\frac{\partial^r
E_0}{\partial [u^{(0)}]^{p_1}\partial [u^{(1)}]^{p_2}\dots
\partial [u^{(k)}]^{p_{k+1}}}\right]_{|_{q=0}}\\\no&&\hspace{2.2cm}\times\prod_{i=1}^n
\left[\sum^{i-1}_{j=0}{i-1\choose
j}\theta^j\widehat{u}^{(0)}_{i-j}\right] ^{p_{i1}}\dots
\left[\sum^{i-1}_{j=0}{i-1\choose
j}\theta^j\widehat{u}^{(k)}_{i-j}\right] ^{p_{i(k+1)}}\\\no
&&\hspace{1.7cm}=\sum_0\sum_1\dots\sum_n\frac{n!}{\prod\limits_{i=1}^n
\prod\limits_{j=1}^{k+1}p_{ij}!}\times\left[\frac{\partial^r
E_0}{\partial [u^{(0)}]^{p_1}\partial [u^{(1)}]^{p_2}\dots
\partial [u^{(k)}]^{p_{k+1}}}\right]_{|_{q=0}}\\\no&&
\hspace{2.2cm}\times[\widehat{u}^{(0)}_{1}]^{p_{11}}\dots[\widehat{u}^{(k)}_{1}]^{p_{1(k+1)}}
[\widehat{u}^{(0)}_{2}+\theta \widehat{u}^{(0)}_{1}
]^{p_{21}}\dots[\widehat{u}^{(k)}_{2}+\theta\widehat{u}^{(k)}_{1}]^{p_{2(k+1)}}\times\dots\\&&
\hspace{2.2cm}\times[\widehat{u}^{(0)}_{n}+\dots+\theta^{n-1}
\widehat{u}^{(0)}_{1}
]^{p_{n1}}\dots[\widehat{u}^{(k)}_{n}+\dots+\theta^{n-1}\widehat{u}^{(k)}_{1}]^{p_{n(k+1)}}.
\end{eqnarray}
In particular,
\begin{eqnarray}\label{fomula}
&&\no\left[\sum^{i-1}_{j=0}{i-1\choose
j}\theta^j\widehat{u}^{(\lambda)}_{i-j}\right]
^{p_{i\nu}}=\sum\frac{p_{i\nu}!\left[\widehat{u}_i^{(\lambda)}\right]^{a_{\nu
0}}}{a_{\nu 0}!\dots a_{\nu (i-1)}!}\left[(i-1)\theta\,
\widehat{u}_{i-1}^{(\lambda)}\right]^{a_{\nu 1}}\dots
\left[\theta^{i-1}\widehat{u}_1^{(\lambda)}\right]^{a_{\nu
(i-1)}}\\\no
&&\hspace{4.3cm}=\sum\frac{p_{i\nu}!\left[\widehat{u}_i^{(\lambda)}\right]^{a_{\nu
0}}}{a_{\nu0}!\dots a_{\nu(i-1)}!}\left[(i-1)
\widehat{u}_{i-1}^{(\lambda)}\right]^{a_{\nu 1}}\dots
\left[\widehat{u}_1^{(\lambda)}\right]^{a_{\nu(i-1)}}\\
&&\hspace{4.8cm}\times \theta^{a_{\nu 1}+2a_{\nu
2}+\dots+(i-1)a_{\nu(i-1)}},
\end{eqnarray}
where $\lambda=0,1,\dots, k$  and the sum is over all nonnegative
integer solutions of the Diophantine equation
$a_{\nu0}+a_{\nu1}+\dots+a_{\nu(i-1)}=p_{i\nu}$.

For Eq.(\ref{total3}), we first consider the terms without
containing parameter $\theta$. That is to say, let $\theta=0$ in
Eq.(\ref{total3}), we have
\begin{eqnarray}\label{total4}
&&\no \frac{\partial^{n} E_0}{\partial q^{n}}_{|_{q=0}}
=\sum_0\sum_1\dots\sum_n\frac{n!}{\prod\limits_{i=1}^n
\prod\limits_{j=1}^{k+1}p_{ij}!}\times\left[\frac{\partial^r
E_0}{\partial [\widehat{u}^{(0)}]^{p_1}\partial
[\widehat{u}^{(1)}]^{p_2}\dots
\partial [\widehat{u}^{(k)}]^{p_{k+1}}}\right]_{|_{q=0}}\\&&\hspace{2.2cm}\times\prod_{i=1}^n
\left[\widehat{u}^{(0)}_{i}\right]
^{p_{i1}}\left[\widehat{u}_i^{(1)}\right] ^{p_{i2}}\dots
\left[\widehat{u}^{(k)}_{i}\right] ^{p_{i(k+1)}},
\end{eqnarray}
which has the same form as Eq.(\ref{total2}).

Secondly, we look for the terms only involving $\theta$ in
Eq.(\ref{total3}). The general formula of this type of term appears
as follows
\begin{eqnarray}\label{total5}
&&\no
\prod_{j=0}^{k}[\widehat{u}^{(j)}_{1}]^{p_{1(j+1)}}\times\dots\times
{p_{m(j+1)}\choose 1}[\widehat{u}^{(j)}_{m}]^{p_{m(j+1)}-1}\theta
\widehat{u}^{(j)}_{m-1}
\times\dots\times[\widehat{u}^{(j)}_{n}]^{p_{n(j+1)}},~~m=2,3,\dots.
\end{eqnarray}
At this time, the parameters satisfy
\begin{eqnarray}\label{para1}
&&\no \sum_0:~
r_1+\dots+(m-1)(r_{m-1}+1)+m(r_m-1)+\dots+n\,r_n=n-1,\\\no&&\sum_{m-1}:~
p_{(m-1)1}+p_{(m-1)2} +\dots+p_{(m-1)(k+1)}=r_{m-1}+1,~~~m=2,\dots,
n,\\\no&&\sum_m:~  p_{m1}+p_{m2}
+\dots+p_{m(k+1)}=r_m-1,~~~m=2,\dots, n,
\\\no&&\sum_{s}:~
p_{s1}+p_{s2} +\dots+p_{s(k+1)}=r_s,~~s=1,\dots,m-2,m+1,\dots, n,
\\\no&&r=p_1+p_2+\dots+p_{k+1}=r_1+r_2+\dots+r_n,\\&&p_j=p_{1j}+p_{2j}+\dots+p_{nj},~~~j=1,2,\dots,k+1.
\end{eqnarray}

Especially, due to $r_1,\dots,r_n$ are nonnegative integers, thus
from the first equation of (\ref{para1}), one has $r_n=0$, which
make condition (\ref{para1}) become
\begin{eqnarray}\label{para2}
&&\no \sum_0:~
r_1+\dots+(m-1)(r_{m-1}+1)+m(r_m-1)+\dots+(n-1)\,r_{n-1}=n-1,\\\no&&\sum_{m-1}:~
p_{(m-1)1}+p_{(m-1)2} +\dots+p_{(m-1)(k+1)}=r_{m-1}+1,~~~~m=2,\dots,
n-1,\\\no&&\sum_m:~  p_{m1}+p_{m2}
+\dots+p_{m(k+1)}=r_m-1,~~~~m=2,\dots, n-1,
\\\no&&\sum_{s}:~
p_{s1}+p_{s2} +\dots+p_{s(k+1)}=r_s,~~s=1,\dots,m-2,m+1,\dots, n,
\\\no&&r=p_1+p_2+\dots+p_{k+1}=r_1+r_2+\dots+r_{n-1},\\&&p_j=p_{1j}+p_{2j}+\dots+p_{(n-1)j},~~~~j=1,2,\dots,k+1.
\end{eqnarray}

The condition (\ref{para2}) is just the requirement of the
parameters in $(\partial^{n-1} E_0/\partial q^{n-1})_{|_{q=0}}$, and
the coefficient of $\theta$ in Eq.(\ref{total3}) corresponds to
$(\partial^{n-1} E_0/\partial q^{n-1})_{|_{q=0}}$ multiplied by
$n(n-1)$. Hence, we prove that Eq.(\ref{total3}) involves
$n(n-1)(\partial^{n-1} E_0/\partial q^{n-1})_{|_{q=0}}$.

Enlarging the degree of $\theta$ and repeating similar procedure as
above up to $\theta^{n-2}$, we find that Eq.(\ref{total3}) contains
$\Big[\sum\limits_{i=2}^{n-1}\frac{n!}{(i-1)!}{n-1\choose
i-2}\theta^{n-i+1}\frac{\partial^{i-1} E_0}{\partial
q^{i-1}}\Big]_{|_{q=0}}$, thus at last, we consider the maximal
degree of $\theta$, that is  $\theta^{n-1}$, which only occurs
 in the terms containing $u_n$ or its derivatives. Hence, this case
 corresponds to $r=1$ in Lemma 1,  which implies 
the terms corresponding to $\theta^{n-1}$ in Eq.(\ref{total3}) reads
\begin{eqnarray}
&&n!\left[\frac{\partial E_0}{\partial u^{(0)}}_{|_{q=0}}u^{(0)}_1+
\frac{\partial E_0}{\partial u^{(1)}}_{|_{q=0}}u^{(1)}_1+\dots+
\frac{\partial E_0}{\partial u^{(k)}}_{|_{q=0}}u^{(k)}_1\right],
\end{eqnarray}
which is just the $(\partial E_0/\partial q)_{|_{q=0}}$ multiplied
by $n!$. This prove it.$~~~~\square$

 Next, for $\theta\neq0$,  we give the main
theorem about the connections between Eq.(\ref{approdeter}) and
Eq.(\ref{homotopydetert}).

 \textbf{Theorem 1.}
Eq.(\ref{homotopydetert}) is equivalent to Eq.(\ref{approdeter})
under the transformation
\begin{eqnarray}\label{transform}
&&
u_0^{(\lambda)}=\widetilde{u}_0^{(\lambda)},~~~~u_l^{(\lambda)}=\sum^{l-1}_{j=0}{l-1\choose
j}\theta^j[\epsilon(1-\theta)]^{l-j}\widetilde{u}_{l-j}^{(\lambda)},
\end{eqnarray}
where $l=1,2,\dots,~\lambda=0,1,\dots,k$.

\emph{Proof.}
 We start with Eq.(\ref{homotopydetert}) and list it for different order of
 $q$ up to $l$ in detail as follows
\begin{eqnarray}\label{expandhomol}
&&\no E_0|_{q=0}=0,\\&&\no \Big[\frac{\partial E_0}{\partial
q}+\epsilon(1-\theta) E_1\Big]_{|_{q=0}}=0,\\\no
&&\Big[\frac{\partial^2 E_0}{\partial q^2}+2\epsilon(1-\theta)
\Big(\frac{\partial E_1}{\partial q}+\theta
E_1\Big)\Big]_{|_{q=0}}=0,\\ &&\Big[\frac{\partial^3 E_0}{\partial
q^3}+3\epsilon(1-\theta) \Big(\frac{\partial^2 E_1}{\partial
q^2}+2\theta\frac{\partial E_1}{\partial q}+2\theta^2
E_1\Big)\Big]_{|_{q=0}}=0,\\\no &&\hspace{5cm}\dots
\\\no&& \Big[\frac{\partial^l E_0}{\partial
q^l}+l\,\epsilon(1-\theta)\Big(\frac{\partial^{l-1} E_1}{\partial
q^{l-1}}+(l-1)\theta\frac{\partial^{l-2} E_1}{\partial
q^{l-2}}+\dots+(l-1)!\,\theta^{l-1} E_1\Big)\Big]_{|_{q=0}}=0.
\end{eqnarray}

The proof of Lemma 1 tells us that $\partial^l E_0/\partial q^l$
with $q=0$ is linear about $u_l\,(l=1,2,\dots)$ and its derivatives,
thus we first consider transformation (\ref{tranY22}) for
Eq.(\ref{expandhomol}).

The first equation in Eq.(\ref{expandhomol}) is unperturbed equation
which is the same as the first equation in Eq.(\ref{approdeter}).
For the second equation, by Lemma 1, $(\partial E_0/\partial
q)_{|_{q=0}}$ is linear about $u_1$ and its derivatives, thus one
can adopt a scaling transformation to convert it to the same form as
the second one in Eq.(\ref{approdeter}).

Next, we consider the third equation in Eq.(\ref{expandhomol}). Take
the first three transformations of (\ref{tranY22}) into the third
equation in Eq.(\ref{expandhomol}) and minus the second equation
multiplied by $2\,\theta$, then by means of Lemma 2, one obtains
\begin{eqnarray}\label{expandhomodetail11}
&& \Big[\frac{\partial^2}{\partial
q^2}E_0(x,\widehat{u}\dots,\widehat{u}^{(k)})+2\epsilon(1-\theta)\frac{\partial}{\partial
q} E_1(x,\widehat{u}\dots,\widehat{u}^{(k)})\Big]_{|_{q=0}}=0,
\end{eqnarray}
which has the same form as the third equation in
Eq.(\ref{approdeter})
\begin{eqnarray}
&&\no\Big[\frac{\partial^2 E_0}{\partial \epsilon^2}+2\frac{\partial
E_1}{\partial \epsilon}\Big]_{|_{\epsilon=0}}=0
\end{eqnarray}
except for the coefficient of $\partial E_1/\partial q$.

In the same way, substituting the first four transformations of
(\ref{tranY22}) into the fourth equation and minus the sum of the
second equation multiplied by $6\,\theta^2$ and
Eq.(\ref{expandhomodetail11}) multiplied by $6\,\theta$, also using
Lemma 2, we have
\begin{eqnarray}\label{expandhomodetail22}
&&\Big[\frac{\partial^3}{\partial
q^3}E_0(x,\widehat{u}\dots,\widehat{u}^{(k)})+3\epsilon(1-\theta)\frac{\partial^2}{\partial
q^2} E_1(x,\widehat{u}\dots,\widehat{u}^{(k)})\Big]_{|_{q=0}}=0,
\end{eqnarray}
which has the same form as the fourth equation in
Eq.(\ref{approdeter})
\begin{eqnarray}
&&\no\Big[\frac{\partial^3 E_0}{\partial \epsilon^3}+3
\frac{\partial^2 E_1}{\partial \epsilon^2}\Big]_{|_{\epsilon=0}}=0
\end{eqnarray}
except for the coefficient of $\partial^2 E_1/\partial q^2$.

Enlarging $l$ inductively and assuming that the first $l$ equations
of Eq.(\ref{expandhomol}) have been transformed to
\begin{eqnarray}\label{expandhomolk}
&&\Big[\frac{\partial^{l-1}}{\partial
q^{l-1}}E_0(x,\widehat{u}\dots,\widehat{u}^{(k)})+\epsilon(l-1)(1-\theta)\frac{\partial^{l-2}}{\partial
q^{l-2}} E_1(x,\widehat{u}\dots,\widehat{u}^{(k)})\Big]_{|_{q=0}}=0,
\end{eqnarray}
then insert the first $(l+1)$ transformations of (\ref{tranY22})
into the $(l+1)$th equation and minus the sum of the $i$th equation
of the form (\ref{expandhomolk})  multiplied by
$\frac{l!}{(i-1)!}{l-1\choose i-2}\theta^{l-i+1}~(i=2,3,\dots,l)$,
we convert the last equation of Eq.(\ref{expandhomol}) to the form
\begin{eqnarray}\label{expandhomolk1}
&&\Big[\frac{\partial^l}{\partial
q^l}E_0(x,\widehat{u}\dots,\widehat{u}^{(k)})+\epsilon
l(1-\theta)\frac{\partial^{l-1}}{\partial q^{l-1}}
E_1(x,\widehat{u}\dots,\widehat{u}^{(k)})\Big]_{|_{q=0}}=0,
\end{eqnarray}
which has the same form as the $(l+1)$th equation in
Eq.(\ref{approdeter})
\begin{eqnarray}
&&\no\Big[\frac{\partial^l E_0}{\partial \epsilon^l}+l\,
\frac{\partial^{l-1} E_1}{\partial
\epsilon^{l-1}}\Big]_{|_{\epsilon=0}}=0
\end{eqnarray}
except for the coefficient of $\partial^{l-1}E_1/\partial q^{l-1}$.

At last, by Lemma 1, using the scaling transformation
\begin{eqnarray}\label{tranY33}
\widehat{u}_0^{(\lambda)}=\widetilde{u}_0^{(\lambda)},~\widehat{u}_1^{(\lambda)}=\epsilon(1-\theta)\widetilde{u}_1^{(\lambda)},
~\widehat{u}_2^{(\lambda)}=\epsilon(1-\theta)\widetilde{u}_2^{(\lambda)},~\dots,~
\widehat{u}_l^{(\lambda)}=[\epsilon(1-\theta)]^l\widetilde{u}_l^{(\lambda)},
\end{eqnarray}
one can convert Eq.(\ref{expandhomolk1}) to the $(l+1)$th equation
of Eq.(\ref{approdeter}) by ASM.

In summary, combining transformations (\ref{tranY22}) and
(\ref{tranY33}), we obtain transformation (\ref{transform}), which
connects Eq.(\ref{approdeter}) and Eq.(\ref{homotopydetert}). This
proves it.$~~~~\square$


Transformation (\ref{transform}) in Theorem 1 build a bridge between
AHSM and ASM for perturbed PDEs, that is to say, there is no need to
perform AHSM to the governing equations because the desired
approximate homotopy series solutions can be obtained by acting the
transformation (\ref{transform}) on the approximate solutions by
ASM.






\section{Applications}
In this section, we apply transformation (\ref{transform}) to study
the connections between ASM and AHSM for the Cahn-Hilliard equation
\cite{cahn,sun}
\begin{eqnarray}\label{rec-h}
u_t=-[F(u)u_x]_x-\epsilon u_{xxxx},
\end{eqnarray}
which is one of the simplest models to characterize the process in
the context of the continuum theory of phase transitions. In
Eq.(\ref{rec-h}), $x,t$ are two independent variables and $u$ is a
scalar dependent one, the nonlinearity $F(u)$ is the derivative of a
double-well potential with wells of equal depth and $0<\epsilon\ll1$
measures the thickness of an interface separating the two  preferred
states of the system. 

\subsection{The connections between the couple equations}
From Eq.(\ref{rec-h}), one has
\begin{eqnarray}
&&\no E_0(u)=u_t+[F(u)u_x]_x,~~~E_1(u)=u_{xxxx}.
\end{eqnarray}
By means of the general formulas of couple system
(\ref{homotopydetert}), we get the  following couple equations by
AHSM
\begin{eqnarray}\label{expandhomo}
&&u_{k,t}+\Big[\sum^{k}_{i=0}\frac{ F^{(j)}(u_0)}{j_0!j_1!\dots
j_i!}u^{j_0}_{l_0}u^{j_1}_{l_1}\dots
u^{j_i}_{l_i}u_{k-i,x}\Big]_x+\epsilon(1-\theta)\sum^{k-1}_{i=0}\theta^{k-1-i}u_{i,xxxx}=0.
\end{eqnarray}
where, hereinafter,
$u_{-1}=0,~u^{j_i}_{l_i}=(u_{l_i})^{j_i},~u_{k,t}=\partial
u_k/\partial t,~u_{i,xxxx}=\partial^4 u_{i}/\partial
x^4,~j_0+j_1+\dots +j_i=j,~l_0j_0+l_1j_1+\dots+ l_ij_i=i,~0\leq
j\leq i$ and $l_0,l_1,\dots, l_i$ are not equal to zero and mutual
inequivalent. $l_i,j_i,j,k~(i,j,k=0,1,2,\dots)$ are nonnegative
integers and satisfy the above relations.

Now, we use transformation (\ref{transform}) to convert
Eq.(\ref{expandhomo}) to the couple equations by ASM. In order to
give specific procedure about how transformation  (\ref{transform})
works for Eq.(\ref{expandhomo}), we first multiply both sides of
Eq.(\ref{expandhomo}) with $k!$ in order to make it  with integer
coefficients  and then list it for different order of $q$ in detail
as follows
\begin{eqnarray}\label{expandhomodetail}
&&\no u_{0,t}+[F(u_0)u_{0,x}]_x=0,\\\no
&&u_{1,t}+[F(u_0)u_{1}]_{xx}+\epsilon(1-\theta) u_{0,xxxx}=0,  \\\no
&&2u_{2,t}+[F'(u_0)u_1^2+2F(u_0)u_2]_{xx}+2\epsilon(1-\theta)
[u_{1,xxxx}+\theta u_{0,xxxx}]=0,\\\no
&&6u_{3,t}+[F''(u_0)u_1^3+6F'(u_0)u_1u_2+6F(u_0)u_3]_{xx}\\
\no&&\hspace{5.25cm}+6\epsilon(1-\theta) [u_{2,xxxx}+ \theta
u_{1,xxxx}+\theta^2u_{0,xxxx}]=0,\\ &&\hspace{5cm}\dots
\\&&\no k!\, u_{k,t}+\Big[\sum^{k}_{i=0}\frac{k!\,
F^{(j)}(u_0)}{j_0!j_1!\dots j_i!}u^{j_0}_{l_0}u^{j_1}_{l_1}\dots
u^{j_i}_{l_i}u_{k-i,x}\Big]_x+\epsilon(1-\theta)\,k!\sum^{k-1}_{i=0}\theta^{k-1-i}u_{i,xxxx}=0.
\end{eqnarray}

Like the proof of Theorem 1, we first consider transformation
\begin{eqnarray}\label{tranY2}
&&
u_0^{(\lambda)}=\widehat{u}_0,~~u_k^{(\lambda)}=\sum^{k}_{i=0}{k-1\choose
i}\theta^i\widehat{u}_{k-i}^{(\lambda)},~~k=1,2,\dots,~
\lambda=0,1,\dots,k.
\end{eqnarray}

By means of Theorem 1, the first two equations of
Eq.(\ref{expandhomodetail}) are equivalent to the corresponding ones
by ASM under a scaling transformation
$u_0=\widetilde{u}_0,u_1=\epsilon(1-\theta)\widetilde{u}_1$, thus we
consider the third equation. Take the first three transformations of
(\ref{tranY2}) into the third equation and minus the second equation
multiplied by $2\theta$, one obtains
\begin{eqnarray}\label{expandhomodetail1}
&&2\widehat{u}_{2,t}+[F'(\widehat{u}_0)\widehat{u}_1^2+2\widehat{u}_2F(\widehat{u}_0)]_{xx}+2\epsilon(1-\theta)\widehat{u}_{1,xxxx}=0.
\end{eqnarray}

Similarly, substitute the first four transformations of
(\ref{tranY2}) into the fourth equation and minus the sum of the
second equation multiplied by $6\theta^2$ and
Eq.(\ref{expandhomodetail1}) multiplied by $6\theta$, we have
\begin{eqnarray}\label{expandhomodetail2}
&&\no6\widehat{u}_{3,t}+[F''(\widehat{u}_0)
\widehat{u}_1^3+6\widehat{u}_2\widehat{u}_1F'(\widehat{u}_0)+6\widehat{u}_3F(\widehat{u}_0)]_{xx}+6\epsilon(1-\theta)\widehat{u}_{2,xxxx}=0.
\end{eqnarray}

Suppose the first $k$ equations in Eq.(\ref{expandhomodetail}) be in
the form
\begin{eqnarray}\label{expandhomodetaik}
&&\no (k-1)!\,u_{k-1,t}+\Big[\sum^{k-1}_{i=0}\frac{
(k-1)!\,F^{(j)}(u_0)}{j_0!j_1!\dots
j_i!}u^{j_0}_{l_0}u^{j_1}_{l_1}\dots
u^{j_i}_{l_i}u_{k-1-i,x}\Big]_x\\&&\hspace{7cm}+\epsilon(1-\theta)(k-1)!\,u_{k-2,xxxx}=0,
\end{eqnarray}
then insert the first $(k+1)$ transformations into the $(k+1)$th
equation and minus the sum of the $i$th equation of the form
(\ref{expandhomodetaik}) multiplied by $\frac{k!}{(i-1)!}{k-1\choose
i-2}\theta^{k-i+1}~(i=2,3,\dots,k)$, we convert the last equation of
Eq.(\ref{expandhomodetail}) to the form
\begin{eqnarray}
\no k!\, u_{k,t}+\Big[\sum^{k}_{i=0}\frac{ k!\,
F^{(j)}(u_0)}{j_0!j_1!\dots j_i!}u^{j_0}_{l_0}u^{j_1}_{l_1}\dots
u^{j_i}_{l_i}u_{k-i,x}\Big]_x+\epsilon(1-\theta)k!\, u_{k-1,xxxx}=0.
\end{eqnarray}

Finally, since the $i$th equation of Eq.(\ref{expandhomodetail}) is
linear about $u_i$ and its derivatives $(i=1,2,\dots,k)$, thus
integrating the scaling transformation
\begin{eqnarray}\label{tranY3}
\no
\widehat{u}_0^{(\lambda)}=\widetilde{u}_0^{(\lambda)},~\widehat{u}_1^{(\lambda)}=\epsilon(1-\theta)\widetilde{u}_1^{(\lambda)},
~\widehat{u}_2^{(\lambda)}=\epsilon(1-\theta)\widetilde{u}_2^{(\lambda)},~\dots,~
\widehat{u}_k^{(\lambda)}=[\epsilon(1-\theta)]^k\widetilde{u}_k^{(\lambda)},
\end{eqnarray}
 and (\ref{tranY2}), one obtains (\ref{transform}), which convert Eq.(\ref{expandhomo})
to the form
\begin{eqnarray}\label{expand}
\no k!\, u_{k,t}+\Big[\sum^{k}_{i=0}\frac{ k!\,
F^{(j)}(u_0)}{j_0!j_1!\dots j_i!}u^{j_0}_{l_0}u^{j_1}_{l_1}\dots
u^{j_i}_{l_i}u_{k-i,x}\Big]_x+k!\, u_{k-1,xxxx}=0,
\end{eqnarray}
 after deleting the symbol $\,\widetilde{}\,$, which is just the couple equations  via formulas (\ref{approdeter}) by ASM.
\subsection{Homotopy series solutions through the transformation}
In this subsection, we use transformation (\ref{transform}) to
obtain approximate homotopy series solutions from the approximate
solutionsby ASM.

\textbf{I. Traveling wave form}

By ASM,  the traveling wave solution with third-order precision for
$F(u)=1/u$ is given by $u=\widetilde{u}_0+\epsilon
\widetilde{u}_1+\epsilon^2
\widetilde{u}_2+\epsilon^3\widetilde{u}_3$ after selecting proper
parameters, where
\begin{eqnarray}\label{solu11}
\no\widetilde{u}_0=\frac{1}{a{\left( a\,t - x \right)
    }},\,\widetilde{u}_1=- \frac{3}{a^2\,{\left( a\,t - x \right)
    }^4},\,\widetilde{u}_2=
    \frac{774}{10\,a^3\,{\left( a\,t - x \right)
    }^7} ,\,\widetilde{u}_3=- \frac{51273}{10\,a^4\,{\left( a\,t - x \right)
    }^{10}},
\end{eqnarray}
and $a$ is wave speed.

By virtue of the first four transformations of (\ref{transform}), i.e.,
\begin{eqnarray}\label{solu12}
&&\no u_0=\widetilde{u}_0,
u_1=\epsilon(1-\theta)\widetilde{u}_1,\\\no&&
u_2=\epsilon(1-\theta)[\epsilon(1-\theta)\widetilde{u}_2+\theta\widetilde{u}_1],\\\no
&&u_3=\epsilon(1-\theta)[\epsilon^2(1-\theta)^2\widetilde{u}_3+2\theta\epsilon(1-\theta)\widetilde{u}_2+\theta^2\widetilde{u}_1],
\end{eqnarray}
we obtain a third-order approximate homotopy series solution of
Eq.(\ref{rec-h}) in the form $u=u_0+q u_1+q^2 u_2+q^3u_3$, which is
\begin{eqnarray}\label{solu2}
&&\no\hspace{-0.6cm} u=\frac{1}{a^2 t-a x}+\frac{3\epsilon q (\theta
-1) }{a^2 (x-a t)^4}+\frac{3\epsilon  q^2(\theta -1) [5 a \theta (a
t-x)^3+129 \epsilon (\theta -1) ]}{5 a^3 (a
t-x)^7}\\&&\hspace{0.1cm} +\frac{3\epsilon  q^3 (\theta-1) [10 a^2
\theta ^2 (x-a t)^6+516 \epsilon \theta  a (\theta -1) (a
t-x)^3+17091 \epsilon ^2(\theta -1)^2 ]}{10 a^4 (x-at)^{10}}.
\end{eqnarray}

 Inserting (\ref{solu2}) into Eq.(\ref{rec-h})
and putting $x=1,t=0.1,\epsilon=0.01$, we get figure 1 which
demonstrates the error of solution (\ref{solu2}) with respect to the
auxiliary parameter $\theta$.
\begin{figure}[htp]
\begin{center}
\includegraphics{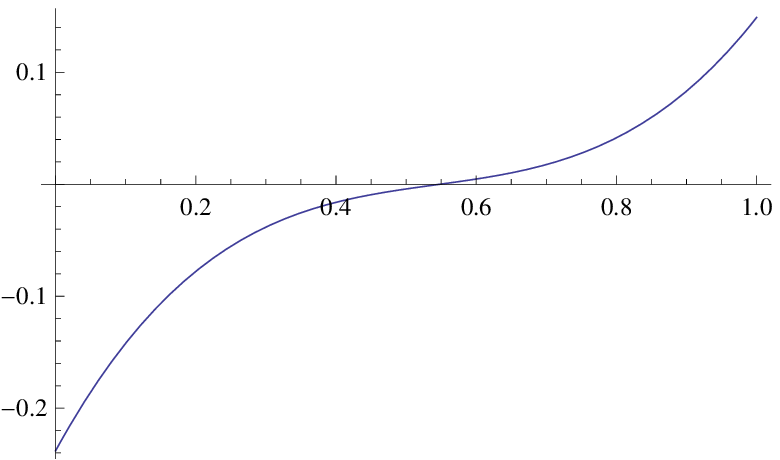}
\end{center}
\caption{{\small $x=1,t=0.1,\epsilon=0.01$, horizontal axis denote
$\theta$,  vertical axis denote the error}}
\end{figure}

 From Figure 1, in particular,
assuming convergence-control parameter $\theta= 0.5478$, one has
\begin{eqnarray}
&&\no u_t+[F(u)u_x]_x+\epsilon u_{xxxx}\approx4.70\times10^{-6}.
\end{eqnarray}
\textbf{II. $F=u$}

We construct the following approximate solutions with fourth-order
precision after assigning proper value to some parameters by ASM
\begin{eqnarray}\label{solu21}
&& \no u=\frac{x^2}{6
   t}+\frac{ \epsilon }{x^2}-\frac{30\epsilon ^2  t }{x^6}-\frac{46440
 \epsilon ^3 t^2}{7 x^{10}}-\frac{804646440 \epsilon ^4 t^3}{203 x^{14}}.
\end{eqnarray}

By virtue of the first five transformations of (\ref{transform}), a fourth-order approximate homotopy series solution follows
\begin{eqnarray}\label{solu4}
&&\no u=\frac{x^2}{6 t}-\frac{ \epsilon q (\theta -1)}{x^2} -\frac{
\epsilon q^2(\theta -1)\left(390 \epsilon  (\theta -1) t+13 \theta
x^4\right)}{13 x^6}\\\no &&\hspace{0.8cm}+\frac{ \epsilon  q^3
(\theta -1)\left(603720 \epsilon ^2(\theta -1)^2 t^2 -5460 \epsilon
\theta (\theta -1) t x^4 -91 \theta ^2 x^8\right)}{91 x^{10}}\\\no
       &&\hspace{0.8cm}+\epsilon  q^4  (\theta -1)\Big[\frac{
       -10460403720  \epsilon ^3(\theta -1)^3 t^3+52523640\epsilon ^2 \theta(\theta -1)^2   t^2 x^4 }{2639 x^{14}}\\
       &&\hspace{0.8cm}-\frac{237510  \epsilon\theta ^2(\theta -1)  t x^8 +2639 \theta ^3 x^{12}}{2639 x^{14}}\Big].
        \end{eqnarray}

When  the  constants  and  variables  in  solution (\ref{solu4}) are
specified by $x=1,t=0.1,\epsilon=0.01$, the relationship between the
error of fourth-order precision approximate homotopy series
solutions of (\ref{solu4}) and parameter $\theta$ is sketched in
Figure 2.
\begin{figure}[htp]
\begin{center}
\includegraphics{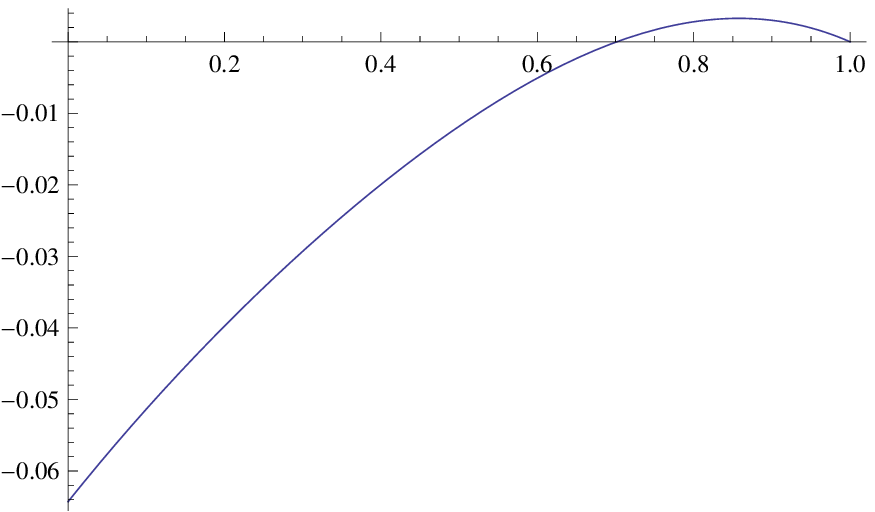}
\end{center}
\caption{{\small $x=1,t=0.1,\epsilon=0.01$, horizontal axis denote
$\theta$,  vertical axis denote the error}}
\end{figure}

Especially, taking convergence-control parameter $\theta=0.7015$,
one has
\begin{eqnarray}
&&\no u_t+[F(u)u_x]_x+\epsilon u_{xxxx}\approx1.59\times10^{-6}.
\end{eqnarray}

\section{Conclusion and discussion}
In this paper, we show that the two couple equations derived by ASM
and AHSM are connected by the transformation (\ref{transform}). As a
result, approximate homotopy series solutions can be obtained by
acting the transformation on the known solutions by ASM. As an
application, the Cahn-Hilliard equation is considered to examine the results.  
This transformation may exert an important role in studying the
perturbed PDEs because it makes researchers facilitate to study the
results by AHSM through the ones by ASM.

In addition, it is worthwhile pointing out that, for
Eq.(\ref{rec-h}), the unique pair of different operators $Y_1,X_1$,
i.e., the approximate homotopy symmetry operator
\begin{eqnarray}
\no Y_1=-\frac{1}{2}x\partial_x
-t\partial_t+\sum^{\infty}\limits_{i=1}\big[i u_i-(i-1)\theta
u_{i-1}\big]\partial_{u_i}
\end{eqnarray}
and the corresponding approximate symmetry one
\begin{eqnarray}
\no X_1=-\frac{1}{2}x\partial_x
-t\partial_t+\sum\limits^{\infty}_{i=0}i
\widetilde{u}_i\partial_{\widetilde{u}_i}
\end{eqnarray}
are equivalent under the transformation
\begin{eqnarray}
&&\no u_0=\widetilde{u}_0,~~~~u_l=\sum^{l-1}_{i=0}{l-1\choose
i}\frac{l-i}{l}\,\theta^i\widetilde{u}_{l-i},~~l=1,2,\dots,
\end{eqnarray}
other than (\ref{transform}). Hence, the relations between the
symmetries of two equivalent equations may form an interesting
topic.

\end{document}